\documentclass[journal]{IEEEtran}
\ifCLASSINFOpdf
\else
\fi

\let\chapter\section
\usepackage{amsmath}
\usepackage{graphicx}
\usepackage{subfigure}
\usepackage{multirow}
\usepackage{amssymb}
\usepackage{algorithmic}
\usepackage{caption}
\usepackage{color}
\usepackage[ruled,lined]{algorithm2e}
\usepackage{bm}
\usepackage{enumerate}
\usepackage{mathrsfs}
\usepackage{leftidx}
\usepackage{authblk}
\usepackage{booktabs}
\usepackage{threeparttable}
\usepackage{array}
\usepackage{amsfonts,amssymb}
\usepackage{hyperref}
\usepackage{textcomp}
\usepackage{cite}

\usepackage{tikz}
\usetikzlibrary{positioning}
\newcommand{\tabincell}[2]{\begin{tabular}{@{}#1@{}}#2\end{tabular}}

\begin{document}
\captionsetup[table]{name={TABLE},labelsep=none}
\captionsetup[figure]{name={Fig.},labelsep=period}

\title{Rapid-flooding Time Synchronization \\for Large-scale Wireless Sensor Networks}

\author{
        Fanrong~Shi,~\IEEEmembership{Student Member,~IEEE},
        Xianguo~Tuo,
        Simon~X.~Yang,~\IEEEmembership{Senior Member,~IEEE},
        \\Jing~Lu
        and~ Huailiang~Li,~\IEEEmembership{Member,~IEEE}

\thanks{Manuscript received September 25, 2018; revised October 16 and May 19, 2019; accepted June 25, 2019. Date of publication XX XX, XXXX; date of current version XX X, XX. This work was supported in part by National Natural Science Foundation of China Programs (61601383), Sichuan Science and Technology Program (No.2018GZ0095), and Longshan academic talent research supporting program of Southwest University of Science and Technology (17LZX650,18LZX633).(Corresponding author: Xianguo Tuo, Huailiang Li.)}

\thanks{Fanrong Shi and Jing Lu are with School of Information Engineering, Southwest University of Science and Technology, Mianyang 621010, China (e-mail: sfr\_swust@swust.edu.cn; lujing\_017@live.cn).}
\thanks{Xianguo Tuo is with Sichuan University of Science and Engineering, Zigong 643000, China (e-mail:tuoxg@cdut.edu.cn).}
\thanks{Simon X. Yang is with Advanced Robotics and Intelligent Systems (ARIS) Laboratory, School of Engineering, University of Guelph, Guelph, Ontario, N1G 2W1, Canada (email: syang@uoguelph.ca). }
\thanks{Huailiang Li is with  the College of Geophysics, Chengdu University of Technology, Chengdu 610059, China. (email: lihl@cdut.edu.cn).}

}

\markboth{}
{Shell \MakeLowercase{\textit{et al.}}: Bare Demo of IEEEtran.cls for IEEE Journals}
\maketitle

\begin{abstract}
Accurate and fast-convergent time synchronization is very important for wireless sensor networks. The flooding time synchronization converges fast, but its transmission delay and by-hop error accumulation seriously reduce the synchronization accuracy.  In this paper, a rapid-flooding multiple one-way broadcast time-synchronization (RMTS) protocol for large-scale wireless sensor networks is proposed. To minimize the by-hop error accumulation, the RMTS uses maximum likelihood estimations for clock skew estimation and clock offset estimation, and quickly shares the estimations among the networks.  As a result, the synchronization error resulting from delays is greatly reduced, while faster convergence and higher-accuracy synchronization is achieved.  Extensive experimental results demonstrate that, even over 24-hops networks, the RMTS is able to build accurate synchronization at the third synchronization period, and moreover, the by-hop error accumulation is slower when the network diameter increases.

\end{abstract}


\IEEEpeerreviewmaketitle

\section{Introduction}
\IEEEPARstart{T}{ime} synchronization is very important to wireless sensor network (WSN) applications, e.g., data acquisition \cite{1,2}, low power \cite{3,TII_7}, location services \cite{4,5}, security \cite{6,TII_6}, networked control \cite{9,10}, industrial WSNs \cite{11, 12},  and smart grid measurement \cite{13, 14}. The traditional network time synchronization protocols, e.g., network time protocol and global positioning system (GPS), may not meet the synchronization requirements in energy-constrained WSN applications resulting from the extra hardware needed or complex protocol employed \cite{16}. The aim of time synchronization algorithms is to correct the local time information on nodes and drive the entire network to obtain a time notion of consistent values.

Low synchronization error, rapid synchronization convergence, and weak-dependency topology management are very important requirements of robust time synchronization in large-scale WSN applications. A faster-convergence approach may adapt rapidly to the changes in clock drifts and network topology, and recover quickly from loss of synchronization.

It is difficult to meet all of the above requirements due to the transmission delay, topology changes, and clock drifts. The RBS \cite{17} and CESP \cite{20} broadcast periodically to build accurate time synchronization among all of the receivers, while fail to meet the synchronization requirements over larger distances \cite{28}. The TPSN \cite{18} is more accurate than the RBS due to less clock offset estimation error on two-way message exchange models, but it is not a distributed approach as topology management is needed to maintain a spanning tree. Average-consensus-based protocols, e.g., GTSP \cite{19}, ATS \cite{21}, CCS \cite{22}, DISTY \cite{TII_5}, and DiStiNCT \cite{23}, are completely independent of topology and more robust to a dynamic WSN, but they cost many more synchronization periods to build the time synchronization. For instance, the synchronization convergence time is up to 120 rounds of synchronization periods in a $5\times7$ grid (diameter of 10) for ATS  \cite{21}. At present, there is no good method to shorten the convergence time for these approaches. The maximum-consensus-based protocols MTS \cite{24} and SMTS \cite{SMTS} converge faster than ATS, but they still cost approximately 90 rounds to converge in a 20-node ring network (diameter of 10), and their convergence time is linear growth of the diameter \cite{24}.

The flooding time synchronization protocols, e.g., FTSP \cite{25}, EGSync \cite{26}, Glossy \cite{27}, PulseSync \cite{28}, FCSA \cite{29}, PISync (FloodPISync and PulsePISync) \cite{PISync}, attracted our attention because they have the potential to be a fast-convergence, accurate, and distributed time synchronization algorithm. The FTSP, FCSA and FloodPISync are slow-flooding protocols, in which the nodes broadcast their local time information periodically and asynchronously, and all of the network nodes synchronize with the root node (reference) when the synchronization is convergent. However, the time information on the root cannot be flooded with the multiple hop nodes quickly, and the accuracy of the time information is reduced by the clock drift on the flooding path. Additionally, the FCSA must maintain the neighbor node table. A rapid-flooding time synchronization protocol (e.g., PulsePISync, PulseSync or Glossy), in which the time information of the root is forwarded to every node as fast as possible, can adapt quickly to changes in clock drift. If the time information is flooded with nodes in a very short time, then there is minimization error on clock drift.

The key idea of flooding time synchronization can be briefly described as follows. All of the nodes synchronize themselves to the reference node that is unique, and the time information of the reference is flooded to the network nodes along multi-hop paths. The synchronization error of adjacent nodes is determined by the time of radio message delivery, which has been carefully analyzed in \cite{26}. The synchronization error of multi-hop nodes depends on the flooding time and flooding path. Thus, the synchronization errors are passed to the next hop node and are accumulated hop by hop. Specifically,

1) the closer the node to the reference node, the higher the node synchronization accuracy \cite{18,26,29}; and

2) the longer the flooding path and the slower the flooding speed, the worse the time information accuracy \cite{18,29}.

Moreover, if a node fails to synchronize to the reference, then all of the nodes on the flooding paths with the failed node will also fail, and even worse. Thus it may take a long time to recover from the damage. Unfortunately, although the time of radio message delivery can be assumed as the Gaussian distribution \cite{30}, it may be very large due to the uncertain software delay. The uncertain delay will be discussed in details in Section II. Therefore, delays remain the major challenge to flooding synchronization approaches in large-scale WSN.

In this paper, we focus on adopting robust and accurate clock parameter estimations to develop the reliable flooding time synchronization for the large-scale WSN. A new rapid-flooding multiple one-way broadcast time synchronization (RMTS) protocol is proposed, which is much more accurate and faster-converging than existing flooding approaches. The multiple one-way broadcast model is proposed to generate time information observations, and only the packet that arrives first will be handled by receivers. Based on the observations, the relative clock skew maximum likelihood estimation (MLE) is used to generate accurate clock skew estimation, and the clock skew estimation sharing is used to guarantee rapid convergence. Furthermore, the relative clock offset MLE is used to guarantee accurate time synchronization, by which the estimation error due to variable delay and uncertain delay is minimized. Even for a 24-hop network (the network capacity reaches $2^{25}-1$ on the binary tree network), the proposed protocol is able to achieve accurate time synchronization at the third round of synchronization periods, i.e., the proposed RMTS takes approximately $3\times$ period time to synchronize all the nodes in the network. The following aspects of RMTS are noteworthy:

1) the clock offset estimation error caused by delay can be minimized, and the RMTS has better time synchronization convergence accuracy than previous approaches;

2) the uncertain delay can be removed from the error link, and the by-hop synchronization error and the probability of adverse effects caused by uncertain delays are significantly reduced; and

3) the convergence rate is significantly improved, and does not depend on the diameter of networks.

The remainder of this paper is organized as follows. We analyze the challenges in flooding time synchronization in Section II and provide the system model in Section III. The RMTS is presented and analyzed in Section IV. Section V provides the implementation and experimental results, in which we also compare the proposed RMTS with FTSP, FCSA, PulseSync and PulsePISync. Finally, some concluding remarks are given in Section VI.

\section{Challenges in Flooding Time Synchronization}
\subsection{Overview}
\begin{figure}[!htb]
\centering
\includegraphics[scale=1.2]{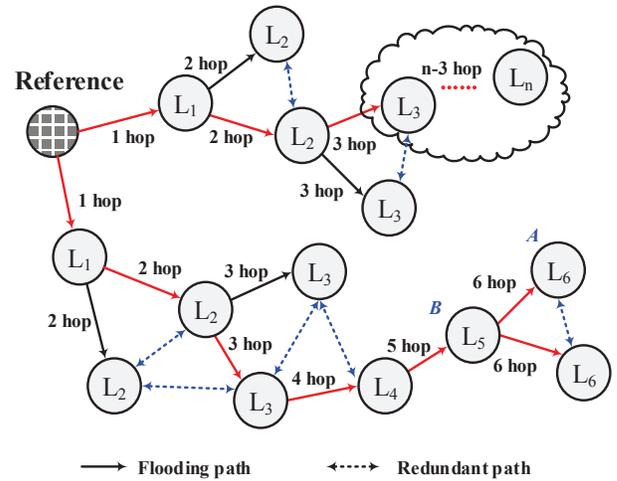}
\caption{Network topology and flood model (diameter of $n$, e.g., $n \geq 6$), in which time information received from the reference is more accurate than that received from closer to the reference.}
\label{fig.1_Flooding}
\end{figure}

The basic flooding model for time synchronization in WSNs is shown in Fig. \ref{fig.1_Flooding}. Considering a large-scale WSN, which always has one or more paths to connect any pair of nodes, the diameter of the WSN is defined as the maximal hop between reference and nodes. The reference node, which can be synchronized with the external clock (e.g., GPS and Coordinated Universal Time), provides the reference clock (global clock) for the networks. The main idea of flooding synchronization is to synchronize every node to the reference, and the basic way is to flood the time information of the reference to the entire network based on one-way broadcast packet transmission.

The experimental results in \cite{29} and \cite{28} show that the synchronization errors of flooding synchronization are mainly caused by delay and clock shift, and they are always accumulating as the hop increases. Nodes that are closer to the reference may receive more accurate time information, e.g., node $B$ receives more accurate time information than $A$.

\vspace{0.4cm}

Therefore, both flooding path and time cost are important to flooding synchronization. In large-scale WSN applications, there are more than one paths between nodes and the flooding time costs are different on each path. Due to the clock shifting, the time information along the path that costs less time is more accurate \cite{26,27,28}. Moreover, the clock drift is always changing, and as a result, the time information is becoming inaccurate, until it is forwarded, and it is likely that the longer the waiting time of time information forwarding, the worse the accuracy.

\subsection{Delay on One-Way Broadcast Model}

A one-way broadcast model is used to collect timestamps in many time synchronization algorithms. The model broadcasts time packets at the real time $t$ to generate pairs of timestamps, e.g., $T_s$ (at the sender) and $T_r$ (at the receiver). If $D$ is the packet transmission delay and $\theta_r^s(t)$ is the relative clock offset, then $T_r-T_s=D+\theta_r^s(t)$. The relative clock offset estimation $\hat{\theta}_r^s(t)$ in one-way-broadcast-based time synchronization is
\begin{equation}\label{equ:1}
  \hat{\theta}_r^s(t)=D+\theta_r^s(t).
\end{equation}

In our previous experiments, we used the Start Frame Delimiter (SFD) interrupt to create MAC-layer timestamps and test $D$ when nodes ran in multi-tasking mode. Three interrupt sources were set for the testbed and three different priority levels configured for the test sequence, i.e., equal priority level for all of interrupt sources and lowest or highest priority level for SFD interrupt. Based on the experimental results (more than 500,000 observations were collected), we found that $D$ on one-way broadcast comprises two portions, i.e.,
\begin{equation}\label{equ:112}
  D=D_{var}+D_{unc},
\end{equation}
where $D_{var}$ is the variable delay and $D_{unc}$ is the uncertain delay. A summary of $D$ is shown in Table \ref{tab:1}.

\vspace{0.7cm}
\begin{table}[htbp]
 \centering
 \captionsetup{justification=centering}
 \caption{\\Summary of delay $D$ on one-way broadcast model. $D_{var}$, variable delay of $D$; $D_{unc}$, uncertain delay of $D$.}{\label{tab:1}}
 \begin{tabular}{cccccc}

  \toprule
  \toprule
         \textbf{SFD}               &  \multicolumn{3}{c}{ \textbf{Variable delay $D_{var}$}}&  \multicolumn{2}{c}{ \textbf{Uncertain delay $D_{unc}$} }  \\
     \tabincell{c}{ \textbf{interrupt}\\ \textbf{priority}}
                          &\tabincell{c}{Probability}
                                               &\tabincell{c}{Mean\\(\textmu s)}
                                                        &STD            &\tabincell{c}{Probability}
                                                                                        &\tabincell{c}{Max.\\(\textmu s)}
                                                                                                                       \\
  \midrule
        Equal	          &0.8825	           &3.322	&0.075	        &0.1175	                &910	         \\
        Lowest	          &0.9463	           &3.330	&0.075	        &0.0537	                &732	           \\
        Highest	          &0.9987	           &3.312	&0.072	        &0.0013	                &910	      \\
  \bottomrule
  \bottomrule
 \end{tabular}
\end{table}
\vspace{-0.3cm}

\begin{figure}[!htb]
\centering
\includegraphics[scale=0.75]{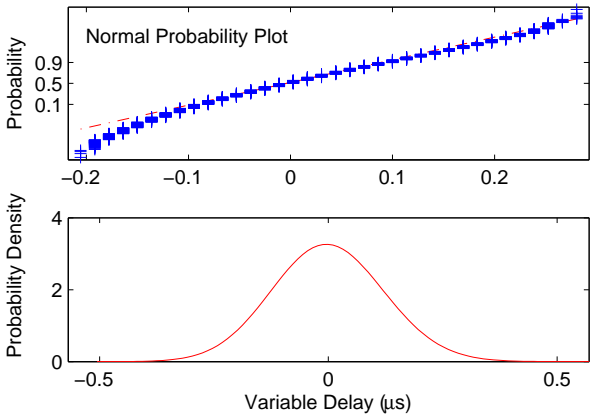}
\caption{Probability and probability density of $d$. Parameter $d$ denotes the variable part of $D_{var}$. The Normal Probability Plot indicates that $d$ is gaussian, and $d\sim N(0,0.0049)$.}
\label{fig.3_delay}
\end{figure}
\vspace{0.4cm}

It is clear that $D_{var}$ is the main portion of $D$, and $D_{var}=D_{fixed}+d$, where $D_{fixed}$ is the fixed portion of $D_{var}$, and $d$ is the variable portion \cite{30}.
The maximum value of $D_{var}$ is approximately 3.6 \textmu s and that of $\hat{D}_{fixed}$ (i.e., the mean of $D_{var}$) is 3.3 \textmu s. After calculating the probability density of variable delay $d$, the results are shown in the bottom part of Fig. \ref{fig.3_delay}. The mean of $d$ is $0$ {\textmu s} and the variance $\sigma^2$ is 0.0049. Using the t-test tool to test $d$, we obtained that $d$ is normal distribution with 99.99\% confidence. The normal probability plot of $d$ is shown in the top part of Fig. \ref{fig.3_delay} and indicates that all of the observations are close to the line and $d$ is Gaussian distributed, i.e., $d\sim N(0,\sigma^2)$. Meanwhile, $D_{unc}$ has no statistical characteristics and is much larger than $D_{var}$, and it depends strictly on the implementation of timestamp, e.g., the software, interrupt nesting, and priority levels.

\subsection{Error Analysis}
We define the synchronization error as $E$. Considering the multi-hop one-way broadcast flooding time synchronization (Synchronization interval of $T_b$), inevitable delay occurs when node floods the reference's time information to its neighbors, which is defined as the flood waiting time $T_{wait}$ $(0<T_{wait}<T_b)$. The clock offset estimation is calculated as in (\ref{equ:1}). Then the synchronization error on a single hop is $E[1]$, and
\begin{equation}\label{equ:17}
 E[1]=D.
\end{equation}

The relative clock frequency speed between $h$-hop and $(h-1)$-hop nodes is $a_h$. Then, the by-hop error accumulation on $k$ hops is
\begin{equation}\label{equ:23}
 E[k]=\sum_{h=0}^{k-1}D[h]+\frac{\sum_{h=0}^{k-1}{(a[h]\times T_{wait}[h])}}{10^6},	
\end{equation}
where $D[h]$ is the delay on the $h$-hop ($D[h]=D_{fixed}+d[h]+D_{unc}[h]$), and $T_{wait}[h]$ is the corresponding forward latency.

The synchronization error of flooding time synchronization (FTSP) can be calculated by (\ref{equ:23}). It is clear that the synchronization error is determined by the delay $D$ and forward latency $T_{wait}$. The rapid-flooding protocols (e.g., Glossy, PulseSync and PulsePISync) minimize $T_{wait}$, and the FCSA and FloodPISync minimize $a[h]$. The FCSA employs the clock speed agreement to maintain $\hat{a}[h]$, and makes all of the nodes at the same speed. Then the synchronization error in FCSA and FloodPISync are given by
\begin{equation}\label{equ:24}
 E_F[h]=\sum_{h=0}^{k-1}D[h]+\frac{\sum_{h=0}^{k-1}{((|a[h]-\hat{a}[h]|)\times T_{wait}[h])}}{10^6}.	
\end{equation}

The rapid-flooding algorithms minimize the forward latency, i.e., $T_{wait}[h]\rightarrow 0$. If the clock skew estimation is accurate enough, i.e., $\hat{a}[h]\rightarrow a[h]$, then the synchronization error in PulsePISync is
\begin{equation}\label{equ:215}
 E_{PI}[h]\thickapprox \sum_{h=0}^{k-1}D[h].	
\end{equation}

Moreover, the PulseSync employs $\hat{D}_{fixed}$ to compensate for the delay $D$, then the synchronization error in PulseSync is
\begin{equation}\label{equ:25}
 E_P[h]\thickapprox \sum_{h=0}^{k-1}(D[h]-\hat{D}_{fixed}).	
\end{equation}

However, the effect of variable $D_{unc}$ is much larger than that of $D_{var}$, and $D_{unc}$ is not considered in previous flooding time synchronization approaches. Specifically, $D_{var}$ is the main source of error for multi-hop time synchronization, which limits the time synchronization convergence accuracy. Meanwhile $D_{unc}$ may cause the entire network to be out of synchronization. According to (\ref{equ:23}), (\ref{equ:24}), (\ref{equ:215}), and (\ref{equ:25}), if a $D_{unc}$ is generated at the intermediate node of the flooding path, then all of the next hop nodes of this flooding path will be out of synchronization, and it will take a very long time to re-synchronize these nodes. According to (\ref{equ:24}) and (\ref{equ:215}), the PulsePISync is more accurate than FloodPISync, thus we will use PulsePISync to compare with RMTS.

\section{System Model}

The WSN in this paper is modeled as the graph $G=(\mathcal{N},\mathcal{E})$, where $\mathcal{N}=\{1,2,\ldots \mathcal{N}\}$ represents the nodes of the WSN and $\mathcal{E}$ defines the available communication links. The set of neighbors for $v_i$ is $\mathcal{N}_i=\{j|(i,j)\in \mathcal{E},i\neq j\}$, where nodes $v_i$ and $v_j$ belong to $\mathcal{N}$, and $j\in \mathcal{N}_i$. There are two time notions defined for the time synchronization algorithm, i.e., the hardware clock $H(t)$ and logical clock $L(t)$.

The hardware clock $H(t)$ is defined as
\begin{equation}\label{equ:3}
  H(t)=\int_0^t h(\tau)d\tau+\theta(t_0),
\end{equation}
where $h(\tau)$ is the hardware clock rate, and it is the inherent attribute of the crystal oscillator and can never be changed or measured. Every node considers itself the ideal clock frequency (nominal frequency), i.e., $h(\tau)=1$. Constant $t_0$ is the moment that a node is powered on and $\theta(t_0)$ is the initial relative clock offset. Therefore, $H(t)$ cannot be changed, and timestamps on $H(t)$ are used to estimate the relative clock speed for the proposed algorithm.

The logical clock $L(t)$ is defined as
\begin{equation}\label{equ:4}
  L(t)=\varphi\times H(t),
\end{equation}
where $\varphi$ is the logical clock rate multiplier and can be changed to speed up or slow down $L(t)$. The timestamps on $L(t)$ are created for clock offset estimation.

Considering the arbitrary nodes $v_i$ and $v_j$, $L_i(t)$ and $L_j(t)$ are the respective logical times, and $\varphi_i^j$ is the relative clock rate, which is
\begin{equation}\label{equ:5}
  \varphi_i^j=1+\theta_\Delta/\tau, (\tau>0),
\end{equation}
where $\theta_\Delta$ is the relative clock offset increment, and $\theta_\Delta=(H_j(t)-H_i(t))-(H_j(t-\tau)-H_i(t-\tau))$.

\section{Proposed RMTS Algorithm}
\subsection{Proposed Multiple One-Way Broadcast}
\begin{figure}[!htb]
\centering
\includegraphics[scale=0.68]{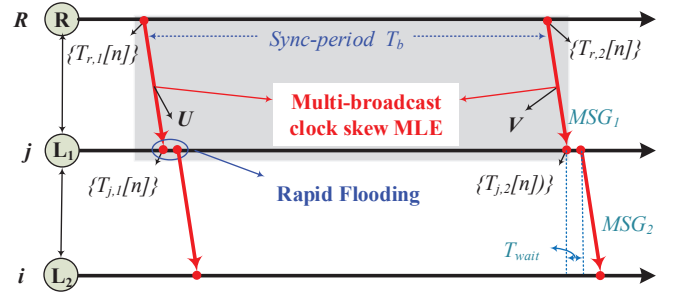}
\caption{Multiple-one-way-broadcast model. Nodes broadcast $N$ times of information packets in a short time interval, then $n$ $(n\leq N)$ pairs of timestamps are generated for the RMTS, e.g., the $T_{r,1}[n]$ and $T_{j,1}[n]$ in process $U$.}
\label{fig.4_broadcast}
\end{figure}

In multiple-one-way-broadcast model, nodes broadcast $N$ time information packets at very short time in a single synchronization period, and $N$ timestamps are generated at both sender and receivers.  As shown in Fig. \ref{fig.4_broadcast}, $\{T_{k,x}[n]\}_{n=1}^N$ is set of timestamps on $v_k$ $(k\in G)$, where $x$ is the identity of the different synchronization process, and $n$ $(n\leq N)$ is the serial number of the one-way broadcast in the same synchronization process.

There are two important processes in the proposed multiple-one-way-broadcast model, i.e., $U$ and $V$, in which an observation set is collected for clock skew estimation and clock offset estimation, e.g., $\langle\{T_{r,1}[n]\}_{n=1}^N,\{T_{j,1}[n]\}_{n=1}^N \rangle$ and $\langle\{T_{r,2}[n]\}_{n=1}^N,\{T_{j,2}[n]\}_{n=1}^N \rangle$, respectively.

\vspace{0.0cm}

\subsection{Clock Skew Estimation}

Here, we introduce MLE of relative clock skew $\hat{\varphi}_j^r$ for $v_r$ and $v_j$, which had been proposed in our previous work and proved to be more accurate than the Linear regression. Based on the down link $U$ and $V$ in Fig. \ref{fig.4_broadcast}, the clock skew estimation observations set $\{u[n],v[n]\}_{n=1}^N$ is calculated by
\begin{equation}\label{equ:6}
  u[n]=T_{j,u}[n]-T_{r,u}[n]=D_u+\theta_u+d_u[n],
\end{equation}
\begin{equation}\label{equ:7}
 v[n]=T_{j,v}[n]-T_{r,v}[n]=D_v+\theta_u+\theta_\Delta+d_v[n],	
\end{equation}
where $D:\{D_u,D_v\}$ are the fixed components of $D_{var}$ and $\{d_u[n],d_v[n]\}_{n=1}^N$ are the variable components of $D$. The parameter $\theta_u$ is the clock offset at $U$ and $\theta_\Delta$is the clock offset increment from $U$ to $V$. It is assumed that $D_u=D_v=D_{fixed}$ \cite{30,32} and $d$ can be modeled by Gaussian distribution.
If $\hat{\theta}_\Delta=v[k]-u[k]$ and $\hat{\tau}=T_{j,v}[k]-T_{j,u}[k]$ $(k\in(0,N))$, then $\hat{\varphi}_j^r$, according to (\ref{equ:5}), is given by
\begin{equation}\label{equ:8}
  \hat{\varphi}_j^r=\frac{\hat{\theta}_\Delta}{T_{j,v}[k]-T_{j,u}[k]}+1,	
\end{equation}
where $T_{j,v}[k]-T_{j,u}[k]\rightarrow T_b$. The estimation errors of $\hat{\varphi}_j^r$ completely depend on $\hat{\theta}_\Delta$, which is given as
\begin{equation}\label{equ:9}
  \hat{\theta}_\Delta=\theta_\Delta+(d_v[k]-d_u[k]).
\end{equation}

We define $P:\{p[k]=\theta_\Delta+(d_v[k]-d_u[k])\}_{k=1}^N.$
Since $d\sim N(\mu,\sigma^2 )$, then the probability distribution function of $\{d_v[k]-d_u[k]\}_{k=1}^N$ is given as
\begin{equation}\label{equ:10}
  f(P)=\frac{1}{2\sqrt{\pi}\sigma}\exp(-\frac{p^2}{4\sigma^2 }).
\end{equation}

The likelihood function for $(\theta_\Delta, \alpha, \sigma^2)$ is given as
\begin{equation}\label{equ:11}
    \begin{aligned}
        L(P;\theta_\Delta,\alpha,& \sigma^2 )\\
                                 & =\frac{1}{(2\sqrt{\pi}\sigma)^N} \exp(-\frac{\sum_{s=1}^N(p-\theta_\Delta)^2}{4\sigma^2 }).
    \end{aligned}
\end{equation}

Differentiating the log-likelihood function yields
\begin{equation}\label{equ:12}
    \frac {\partial \ln L(P;\theta_\Delta,\alpha,\sigma^2 )}{\partial \theta_\Delta}=\frac{1}{2\sigma^2} \sum_{s=1}^N(p-\theta_\Delta).
\end{equation}

Therefore, the MLE of $\hat{\theta}_\Delta$ is given as
\begin{equation}\label{equ:13}
     \begin{aligned}
        \hat{\theta}_{\Delta(MLE)} &=\arg{ \max{[\ln L(P;\theta_\Delta,\alpha,\sigma^2 )]} }\\
                                   &=\frac{\sum_{s=1}^Np}{N}=\bar{P}.
    \end{aligned}
\end{equation}

The MLE of $\hat{\varphi}_j^r$ with the Gaussian delay model is given as
\begin{equation}\label{equ:14}
 \hat{\varphi}_{j(MLE)}^r=\frac{\bar{P}}{\hat{\tau}},
\end{equation}
where $\hat{\varphi}_{j(MLE)}^r$ is independent of clock offset estimation.

\subsection{Clock Offset Estimation}

Rewriting (\ref{equ:1}), the clock offset estimation error on a single hop is calculated as $D$. The MLE of clock offset estimation is simply calculated based on the multiple-one-way- broadcast model and the observations in (\ref{equ:6}) (i.e., $\{u[n]\}_{n=1}^N$), and is given by
\begin{equation}\label{equ:15}
  \hat{\theta}_{i(MLE)}^j=\min_{1\leq n \leq N}{\{u[n]\}_{n=1}^N},	
\end{equation}
where the estimate error is $D_{fixed}+\min(\{d_u[n]\}_{n=1}^N), d\sim N(0,\sigma^2)$. When using the statistical average of delays during a calibration phase \cite{28,29}, i.e., $\hat{D}_{fixed}$ is rewritten as
\begin{equation}\label{equ:16}
 \hat{\theta}_{i(MLE)}^j=\min_{1\leq n \leq N}{\{u[n]\}_{n=1}^N}-\hat{D}_{fixed}.	
\end{equation}
The clock offset estimate error is then $\min{(\{d-u[n]\}_{n=1}^N)}$.

\subsection{Proposed RMTS}

A root (reference) algorithm and non-root algorithm are designed in the RMTS protocol.

\emph{\textbf{1) Root algorithm}}

The root only broadcasts the time information packets and never handles the received time information packets. The pseudo-code of the root algorithm is presented in Algorithm 1.
The $\varphi_R$ of the root is a constant value and can never be changed, i.e., $\varphi_R=1$ (Algorithm 1, Line \ref{alg1.2}). If there is an external clock $L_{ExtRef}$ for the root, such as GPS, the root can synchronize its logical time $L_R$ to $L_{ExtRef}$ (Algorithm 1, Lines \ref{alg1.4} and \ref{alg1.5}). Then, the logical time of the root is updated by $L_R(t+\tau)=(H_R(t+\tau)-H_R(t))+L_{ExtRef}(t),\tau\geq0$; else, $L_R(t)=H_R(t)$.

The root broadcasts periodically to distribute the reference time information packets to neighbors, as in Algorithm 1, Lines \ref{alg1.6}-\ref{alg1.11}. Once the broadcast task is triggered, the root rapidly broadcasts $N$ packets in a very short time interval (there is a fixed clock offset for each broadcast), as in Algorithm 1, Lines \ref{alg1.7}-\ref{alg1.9}. Two groups of timestamps are created over the phase of broadcasting, i.e., timestamp $H_R[n]$ (created on $H_R(t)$) and timestamp $L_R[n]$ (created on $L_R(t)$). The basic information of the broadcast packets comprises four parts: $H_R[n]$, $L_R[n]$, $\varphi_R$, and $ID_R$ (Algorithm 1, Line \ref{alg1.8}).

\emph{\textbf{2) Non-root algorithm}}

The pseudo-code of the non-root algorithm is presented in Algorithm 2. Node $v_i$ is synchronized to the root by calibrating $L_i(t)$ based on the rate multiplier $\varphi_i$ and clock offset estimation $\hat{\theta}_{i(MLE)}^j$, where $v_j$ is the neighbor of $v_i$ and is closer to the root.

\LinesNumbered
\begin{algorithm}[ht]
\caption{{Root algorithm pseudo-code for RMTS. $v_R$ is the root node and $N$ is the maximal number of multiple broadcasts.}}
\textbf{Initialization:}                                {\label{alg1.1}}\\
\quad{Set $\varphi_R=1$, $ID_R=1$,  $n=0$}              {\label{alg1.2}}\\
\quad{Start periodic broadcast task, period of $T_b$}        {\label{alg1.3}}\\
\BlankLine
\If {external clock is used}{                           {\label{alg1.4}}
      {Set $L_R(t)\leftarrow L_{ExtRef}(t)$}\\            {\label{alg1.5}}
    }
\BlankLine
\if\textbf{Upon triggering of broadcast task:}{\\         {\label{alg1.6}}
  \eIf{($n<N$)} {                                        {\label{alg1.7}}
         Broadcast $\langle H_R[n],L_R[n],\varphi_R,ID_R\rangle $\\ {\label{alg1.8}}
         Set $n=n+1$, back to (if ($n<N$))                          {\label{alg1.9}}
    }{                                                              {\label{alg1.29}}
         $ID_R= ID_R+1$, $n=0$;                                     {\label{alg1.10}}
    }                                                               {\label{alg1.11}}
}
\end{algorithm}

\LinesNumbered
\begin{algorithm}[ht]
\caption{{Non-root algorithm Pseudo-code for RMTS. $v_i$ is non-root node, $j\in N_i, (i,j)\in G$.}}
\textbf{Initialization}\;                                                                               {\label{alg2.1}}
      \quad{  Set $\hat{\varphi}_{i(MLE)}^j=1$, $\varphi_i=1$}\\                                   {\label{alg2.2}}
      \qquad{   $\hat{\theta}_{i(MLE)}^j=0$, $n=0$, $ID_i=0$   }\\                                       {\label{alg2.3}}
      \qquad{   $\langle\{H_i[n],H_j[n]\}_{n=1}^N \rangle_{old}\leftarrow 0 $        }\\            {\label{alg2.4}}
      \qquad{   $\langle\{H_i[n],H_j[n]\}_{n=1}^N \rangle_{new}\leftarrow 0  $  }\\                 {\label{alg2.5}}
      \qquad{   $\langle\{L_i[n],L_j[n]\}_{n=1}^N \rangle\leftarrow 0         $    }\\              {\label{alg2.6}}
\BlankLine

\if\textbf{Once $\langle H_j[n],L_j[n], \varphi_j, ID_j\rangle$ is received:}\\                         {\label{alg2.7}}
{  \If{($ID_j>ID_i$)} {                                                                                 {\label{alg2.8}}
        Store $\varphi_j$, $\langle H_i[n], H_j[n]\rangle_{new}$ \\                                      {\label{alg2.9}}
        Store $\langle L_i[n], L_j[n]\rangle$ \\                                                         {\label{alg2.10}}
        Update $ID_i\leftarrow ID_j$ \\                                                                 {\label{alg2.11}}
        Start parameter estimation compensation task \\                                                 {\label{alg2.12}}
    }                                                                                                   {\label{alg2.13}}
}                                                                                                       {\label{alg2.14}}
\BlankLine
\if\textbf{Upon triggering of compensation task:}{\\                                                           {\label{alg2.15}}
    \quad{Calculate $\hat{\varphi}_{i(MLE)}^j$, $\varphi_i\leftarrow\hat{\varphi}_{i(MLE)}^j\times\varphi_j$}\\    {\label{alg2.16}}
    \quad{Calculate $\hat{\theta}_{i(MLE)}^j$}                                                            \\                {\label{alg2.17}}
    \quad{Logical clock compensation}\\                                                                                        {\label{alg2.18}}
    \quad{$\langle\{H_i[n], H_j[n]\}_{n=1}^N\rangle_{old}\leftarrow\langle\{H_i[n], H_j[n]\}_{n=1}^N\rangle_{new}$ }\\   {\label{alg2.19}}
    \eIf {($n<N$) (Rapid flooding)}{                                                                                             {\label{alg2.20}}
        Broadcast $\langle H_i[n], L_i[n], \varphi_i, ID_i\rangle$ \\                                                             {\label{alg2.21}}
        Set $n=n+1$, back to (if ($n<N$))\\                                                                                         {\label{alg2.22}}
    }{                                                                                                                             {\label{alg2.23}}
     $n=0$ \\                                                                                                                       {\label{alg2.24}}
     }                                                                                                                               {\label{alg2.25}}
}                                                                                                                                      {\label{alg2.26}}
\end{algorithm}                                                                                                                         {\label{alg2.27}}

\emph{\textbf{Rapid flooding:}} An identification number $ID$ is used to mark the multiple broadcast packets for rapid flooding. Parameter $ID_R$ is initialized as 1 (Algorithm 1, Line \ref{alg1.2}) and incremented at $v_R$ when a broadcast task is finished (Algorithm 1, Line \ref{alg1.10}). The packets of the current broadcast task are embedded in the same $ID_R$. The received $ID_j$ and local $ID_i$ are used to handle the redundant information packets, which may arrive from different paths (Algorithm 2, Line \ref{alg2.8}).

If the packets from $v_j$ have been handled, then $ID_i$ will be set as $ID_j$, and the following packets that are embedded with the same identify number will be ignored. Although there are multiple flooding paths between the reference and arbitrary node, only one path (the shortest or the fastest) in RMTS (FTSP, PulseSync and PulsePISync) is valid for the time synchronization message. In other words, flooding time synchronization protocols convert the complex network into a set of lines, thus synchronization of complex networks is simplified to the synchronization process of lines.

Once $v_i$ has handled the received time information packets, the compensation task will be triggered. Parameters $\varphi_i$ and $\hat{\theta}_{i(MLE)}^j$ are updated and used to compensate the local logical clock, and timestamps $\langle\{H_i[n], H_j[n]\}_{n=1}^N\rangle_{new}$ will be saved as $\langle\{H_i[n], H_j[n]\}_{n=1}^N\rangle_{old}$ for the next period of synchronization. After a very short random delay (to avoid possible collisions), node $v_i$ will forward the time information and share the clock parameters (i.e., $\varphi_i$). The basic time information of the broadcast packets also comprises four parts, $H_i[n], L_i[n], \varphi_i$, and $ID_i$.

\emph{\textbf{Relative clock offset estimation:}} The parameter $\hat{\theta}_i$ is calculated by (\ref{equ:16}), as in Algorithm 2, Line \ref{alg2.17}. The logical timestamps $\{L_i[n], L_j[n]\}_{n=1}^N$ are used, i.e., $\hat{\theta}_{i(MLE)}^j=\min\limits_{1\leq n\leq N}\{L_i[n]-L_j[n]\}_{n=1}^N-\hat{D}_{fixed}$.

\emph{\textbf{Relative clock skew estimation}}: The rate multiplier $\varphi_i$ of the non-root $v_i$ is initialized as 1 (Algorithm 2, Line \ref{alg2.2}), and will be shared with the neighbors as a part of the time synchronization information packet (Algorithm 2, Line \ref{alg2.21}). For the purpose that the $L_i(t)$ runs at the same rate as $L_R(t)$, $\varphi_i$ is updated by multiplying $\hat{\varphi}_{i(MLE)}^j$ by $\varphi_j$, as in Algorithm 2, Line \ref{alg2.16}. The $\hat{\varphi}_{i(MLE)}^j$ is the MLE of the relative hardware frequency rate and is calculated by (\ref{equ:14}). Here, the hardware clock timestamps are used to calculate it, i.e., $u[n]=\langle H_i[n]-H_j[n]\rangle_{old}$ in (\ref{equ:6}) and $v[n]=\langle H_i[n]-H_j[n]\rangle_{new}$ in (\ref{equ:7}).

\emph{\textbf{Logical clock maintenance}}: After $\varphi_i$ and $\hat{\theta}_{i(MLE)}^j$ are calculated and the local logical clock has been compensated, then the logical time will be updated by
$L_i(t+\tau)=L_i(t)+\hat{\theta}_{i(MLE)}^j+(H_i(t+\tau)-H_i(t))\times\varphi_i$ $(\tau>0)$.

\emph{\textbf{Root election}}: The specified node is set as the root at the initialization phase of the WSN. When the root fails, RMTS will elect a new root. If the node that is closest to the failed root does not receive the time synchronization broadcast for the duration of several (e.g., 2) consecutive synchronization rounds, then it will switch to a new root immediately. Since a node only handles the first arriving message with the largest $ID$, it is easy to elect a unique root. Such a root election approach is also used in the FTSP, FCSA, and PulseSync.

\emph{\textbf{Synchronization error}}: Based on (\ref{equ:16}) and (\ref{equ:25}), the uncertain delay $D_{unc}$ could be remove from the error link of the RMTS. Then the synchronization error of the RMTS is given by
\begin{equation}\label{equ:109}
 E_R[k]\thickapprox \sum_{h=0}^{k-1}(D[h]-\hat{D}_{fixed}-D_{unc}[h])\thickapprox \sum_{h=0}^{k-1}d[h].	
\end{equation}

\section{Testbed Experiments and Discussions}

A prototype system was built around Synchronous Sensing Wireless Sensor (\emph{SSWS}) nodes, whose design was based on CC2530. An external 32 MHz crystal oscillator was set as the system clock source of (\emph{SSWS}), and 5V DC power supplies were used to power the nodes. The synchronization algorithms were implemented in the IEEE802.15.4 Medium Access control (MAC) software stack TIMAC \cite{timac}, and the transmission power was programmed as 0 dBm. The MAC Timer 2 (with a 16-bit timer and a 24-bit overflow counter) of CC2530 worked on up mode and was employed to generate the clock notion. In other words, the clock granularity is 1 {\textmu s} for the time notion, and we cannot distinguish synchronization errors that are smaller than 1 {\textmu s}.

\subsection{Testbed setup}
We use an indoor testbed of 25 \emph{SSWSs} and a sink-node, which were placed in close proximity. Since all of the nodes used the same channel and were all in the communication range of each other, random delay flooding method is adopted in software, and the CC2530's radio is configured as the slotted CSMA-CA transmit mode to avoid the possible collisions. The linear regression table size is eight in FTSP, FCSA, and PulseSync. The multiple number is five in RMTS. The synchronization interval is 30 s. The prior delay $\hat{D}_{fixed}$ is set as 3 {\textmu s} in RMTS, PulseSync, FCSA and PulsePISync. Although \cite{29} does not mention compensating for clock offset estimation by  $\hat{D}_{fixed}$, we do this to make the experimental results of PulsePISync and FCSA more accurate. The results reported are based on measurement experiments conducted over more than 4 h.

\emph{\textbf{1) SFD based MAC-Layer timestamp}}

the CC2530's radio allows software to read the local clock when transmitting or receiving SFD bytes, and embed bytes to the MAC payload when packet is sending. The SFD interrupt request will be triggered once the SFD byte has been transmitted or received by radio, and the MAC-Layer timestamp can be created by the interrupt service function or the timer 2 compare function. The main software delay of timestamp is the interrupt handling time and timestamp response time. The difference of delay between sender and receiver present typical statistical characteristics (reported in Section II.B), and it depends on the processing rate, bit wide of processor, the software efficiency and interrupt priority level.

\emph{\textbf{2)Synchronization error measurement}}

similar to \cite{25,28,29}, the sink-node, which is connected to a PC by serial port, is employed to transmit the testing packet periodically at the interval of 10 seconds; once test node receives a packet, it creates a timestamp immediately and upload the timestamp packets to the PC via the sink-node. The synchronization error is calculated on the PC by pairs of timestamps.

\emph{\textbf{3)Measurement error}}

base on STM32 platform and two \emph{SSWSs}, the testbed is built to test the measurement error. The timer of STM32 is set as two channel input capture model, and its clock source is set as 72 MHz. The timer captures the rising edges of the output signals and records the corresponding count values. Moreover, the tesbed could be used to measure the delay that has been discussed in Section II.B. More than 500,000 data points were collected. The mean of measurement error is about 0.07 {\textmu s}; the max value is about 0.4 {\textmu s}; the variance is 0.0033. Hence, once the clock granularity for the testbed time notion is greater than the max value of measurement error, then the measuring error is in the tolerable range.

\subsection{Experimental results on line network}
We enforce a line topology in software, i.e., \textbf{R}$\leftarrow$\textbf{1}$\dots\leftarrow$\textbf{24} (diameter of 24), in which the \emph{SSWSs} can only communicate with neighbors and sink-node.

\emph{\textbf{1) Local synchronization error}}

\begin{figure}[!htb]
\centering
\includegraphics[scale=0.75]{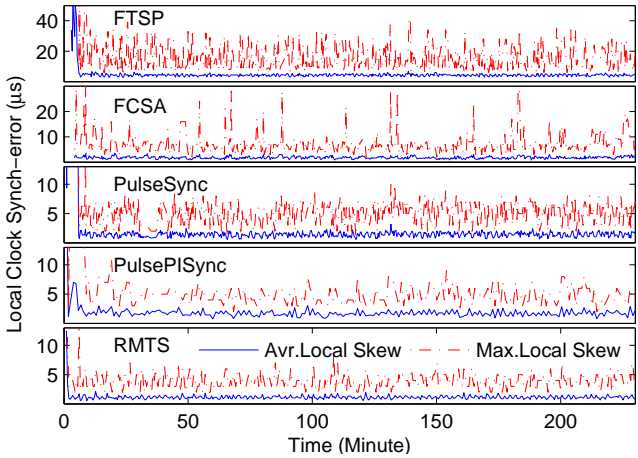}
\caption{Local synchronization error on line topology (24 hops) using the FTSP, FCSA, PulseSync, PulsePISync and RMTS.}
\label{fig.5_LocalSkew}
\end{figure}

\begin{figure}[!htb]
\centering
\includegraphics[scale=0.78]{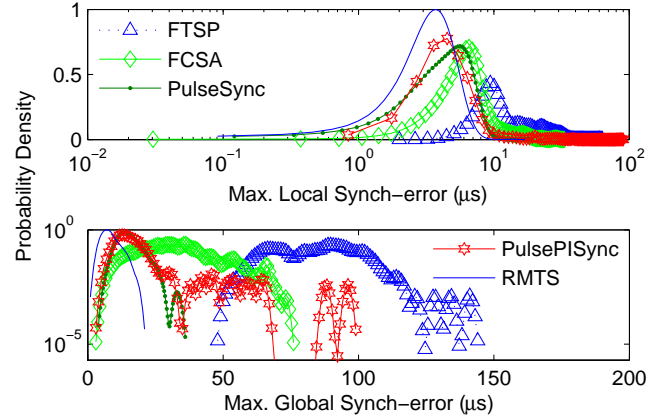}
\caption{Probability density of maximal local synchronization error and maximal global synchronization error. To show more details, logarithmic coordinates at X-axis and Y-axis are used for the local synchronization error and global synchronization error, respectively.}
\label{fig.9_Local_PDF}
\end{figure}

\begin{figure}[!htb]
\centering
\includegraphics[scale=0.78]{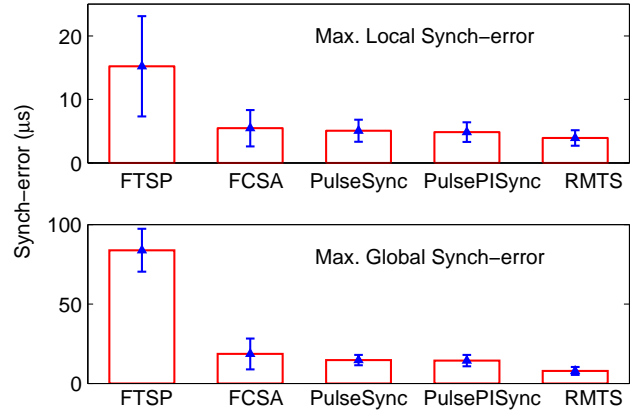}
\caption{Error bars of maximal local synchronization error and maximal global synchronization error. The text annotation indicates are mean and standard deviation.}
\label{fig.12_ErroBars}
\end{figure}

Local synchronization error is employed to evaluate the synchronization error between any two adjacent nodes. The results shown in Fig. \ref{fig.5_LocalSkew} indicate the instantaneous average and instantaneous maximum of local synchronization error. The probability density of the maximal local synchronization error is shown in the upper panel of Fig. \ref{fig.9_Local_PDF}. The time-average and standard deviation of maximal local synchronization error is shown in the upper panel of Fig. \ref{fig.12_ErroBars}.

\vspace{0.3cm}

\emph{i.)} Considering FTSP, $D_{fixed}$ is introduced into the local synchronization error directly, and the probability density of the maximal local synchronization error is loose and its peak is far away from zero. The maximal local synchronization error is up to 15 {\textmu s}.

\emph{ii.)}  In the FCSA, PulseSync and PulsePISync, which use $\hat{D}_{fixed}$ to compensate the clock offset estimation, the local synchronization error is much lower, and the probability density of the maximal local synchronization error is tighter and its peaks is close to zero. The performances of PulseSync and PulsePISync are slightly better than that of FCSA and FTSP with rapid flooding;

\emph{iii.)}  The RMTS further improves the local synchronization error of rapid flooding, and more accurate local synchronization is achieved using clock skew estimation and clock offset estimation. The RMTS is more efficient in optimizing the clock offset estimation error and local synchronization error.

The 95\% confidence interval of the maximal local synchronization error is 3.82-4.01 {\textmu s} in RMTS, while it is 4.70-4.96 {\textmu s} in PulsePISync, 4.96-5.15 {\textmu s} in PulseSync, 7.28-7.65 {\textmu s} in FCSA, and 14.75-15.45 {\textmu s} in FTSP.

\vspace{0.3cm}

\vspace{0.3cm}

\emph{\textbf{2) Global synchronization error}}

\begin{figure}[!htb]
\centering
\includegraphics[scale=0.75]{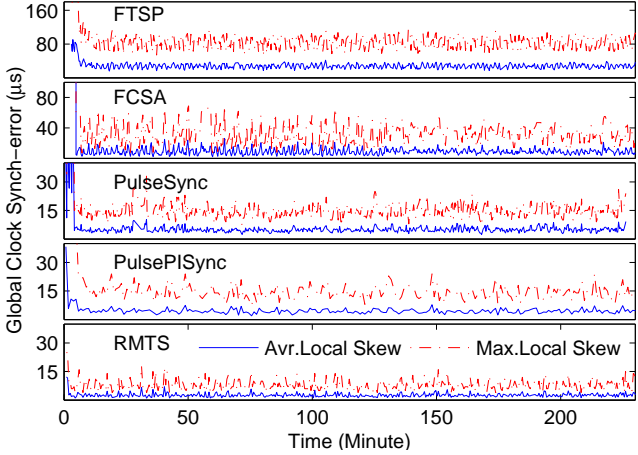}
\caption{Global synchronization error on line topology using the FTSP, FCSA, PulseSync, PulsePISync and RMTS.}
\label{fig.6_GlobakSkew}
\end{figure}

Global synchronization error is employed to evaluate the synchronization error of arbitrary pairs of nodes. Figure \ref{fig.6_GlobakSkew} indicates the instantaneous average and instantaneous maximum of global synchronization error. The probability density of the maximal global synchronization error is shown in the lower panel of Fig. \ref{fig.9_Local_PDF}. The time average and standard deviation of maximal global synchronization error is shown in the lower panel of Fig. \ref{fig.12_ErroBars}.

The experimental results indicate that the probability density of the maximal global synchronization error in RMTS is much tighter and closer to zero. The 95\% confidence interval of the maximal global synchronization error is 7.7-8.11 {\textmu s} in RMTS, while it is 14.04-14.65 {\textmu s} in PulsePISync, 14.52-14.44 {\textmu s} in PulseSync, 18-18.9 {\textmu s} in FCSA, and 83.04-84.26 {\textmu s} in FTSP.

\vspace{0.3cm}

\emph{\textbf{3) Synchronization error to root}}

\begin{figure}[!htb]
\centering
\includegraphics[scale=0.75]{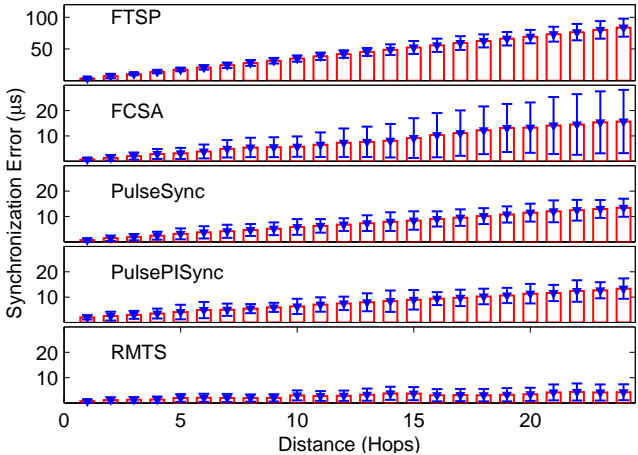}
\caption{Time average of maximal synchronization error from reference. Bars indicate standard deviation over time.}
\label{fig.13_FTSP_FCSA_HOP}
\end{figure}

The synchronization error to root is used to indicate the by-hop error accumulation when the time synchronization is convergent, which is very important to time synchronization in large-scale WSNs. The synchronization error to root commonly increases as the diameter increases. A slower growth speed is better and guarantees smaller synchronization error to root. The time average of maximal synchronization error from the reference is shown in Fig. \ref{fig.13_FTSP_FCSA_HOP}. Both the mean and standard deviations are smaller and the error growth speed in RMTS is slower than the compared methods.

\vspace{0.3cm}

\emph{\textbf{4) Convergence time}}

The convergence times of the experimental results that are shown in Fig. \ref{fig.6_GlobakSkew}, are approximately 6-7 min in FTSP and FCSA, approximately 4 min in PulseSync and PulsePISync, and less than 2 min in RMTS. The shortening to 3 rounds of synchronization periods in RMTS results from the use of the proposed MLE of clock skew and clock parameter sharing. For approximately 50 rounds of unreported experimental results, the convergence time is approximately 10-24 round of synchronization period in FTSP and FCSA, approximately 8-10 round in PulseSync and PulsePISync, and approximately 2-6 round in RMTS. The convergence speed in RMTS is faster, and the RMTS has the potential to meet the challenges of root failure and re-election, topology change, and uncertain delay.

\subsection{Experimental results on grid network}
\begin{figure}[!htb]
\centering
\includegraphics[scale=0.75]{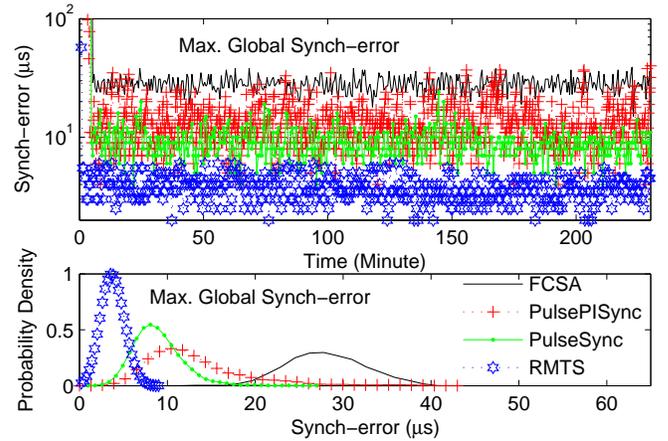}
\caption{Maximal global synchronization error on $5\times5$ grid (diameter of 8) topology using the RMTS, PulseSync, FCSA, and PulsePISync. To show more details, logarithmic coordinates are used for the Y-axis label in upper panel of figure. The 95\% confidence interval of the mean is 27.63-28.47 {\textmu s} in FCSA, 8.69-9.17 {\textmu s} in PulseSync, 12.86-13.41 {\textmu s} in PulsePISync, 3.69-3.87 {\textmu s} in RMTS.}
\label{fig.13_Grid}
\end{figure}

To evaluate the performances of RMTS in the complex network, we enforce a $5\times5$ grid (diameter of 8) topology in software, in which the one in the upper left corner is the reference node, and the \emph{SSWSs} can only communicate with neighbors and sink-node.The experimental results are shown in Fig. \ref{fig.13_Grid}.

As discussed in Section IV.D, the flooding time synchronization protocols convert the complex network into lines. Therefore, the diameter of lines can be considered to be the main factor determining the synchronization error, meanwhile the shape of network affects the length (diameter) of lines. Moreover, the flooding speed will be slow down to avoid the possible collision and the synchronization error will be increased in the complex network (especially for high-density mesh networks).

The time average and standard deviation of the maximal global synchronization error is approximately 3.78 {\textmu s} and 0.9166 in RMTS, respectively, while they are 13.14 {\textmu s} and 5.6681 in PulsePISync, 8.93 {\textmu s} and 2.6589 in PulseSync, 28.05 {\textmu s} and 4.3575 in FCSA.

\vspace{0.3cm}

\subsection{Synchronization period, accuracy, and energy efficiency}
The re-synchronization is required to hold on accurate synchronization, but frequent re-synchronization will lead to high synchronization costs in communication and energy. In energy-constrained WSN applications, it is very important to balance synchronization cost and accuracy. Only communication cost is discussed here because communication is the main part of energy consumption in WSNs.

\begin{figure}[!htb]
\centering
\includegraphics[scale=0.72]{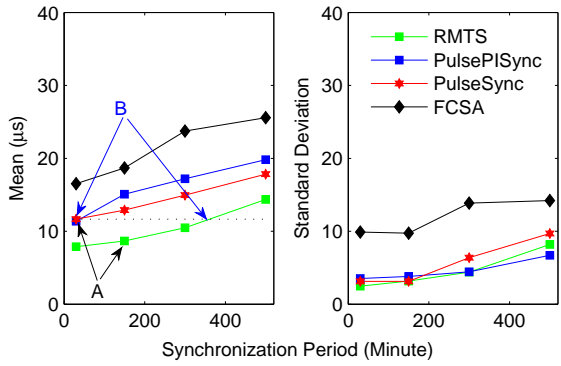}
\caption{Time average and standard deviation of maximal global synchronization error on synchronization intervals of 30, 150, 300, and 500 s. The RMTS has the same communication cost as the PulseSync and PulsePISync at A and the equivalent synchronization error as the PulseSync and PulsePISync at B.}
\label{fig.17_Mean}
\end{figure}

The synchronization periods in the experiments were set as 30, 150, 300, and 500 s. The experiments ran for approximately 4-7 h for RMTS, FCSA, PulseSync and PulsePISync at diverse synchronization period. The maximal synchronization errors were calculated, and their statistical averages (means) and standard deviations are shown in Fig. \ref{fig.17_Mean}. The PulseSync, PulsePISync and RMTS perform better than FCSA.

\vspace{0.3cm}

Considering the communication costs on the premise of equivalent synchronization accuracy, as demonstrated by points B in Fig. \ref{fig.17_Mean}, PulseSync, PulsePISync and RMTS have equivalent synchronization accuracy, i.e., equal to the mean and similar to the standard deviation. The synchronization period of PulseSync and PulsePISync is 30 s, and the communication cost is $1\times120$ broadcast per hour, while the synchronization period of RMTS is approximately 360 s and its communication cost is $5\times10$ broadcast per hour.

Considering the synchronization accuracy on the premise of the same communication cost, i.e., $1\times120$ broadcast per hour, as shown by points A in Fig. \ref{fig.17_Mean}, the means and standard deviations are approximately 7.9 {\textmu s} and 2.48 in RMTS and approximately 11.7 {\textmu s} and 3.12 in PulseSync.

Hence, the energy efficiency of the RMTS is better than that of PulseSync with equivalent synchronization accuracy, and the RMTS is more accurate under the same energy efficiency. Moreover, the synchronization accuracy does not decrease too much even in the 500 s synchronization interval, i.e., the mean is approximately 14.37 {\textmu s} and the standard deviation approximately 8.19. Considering the root failure, the RMTS requires less than 7 rounds (it is 420 seconds in a 30 s synchronization interval) of synchronization period to elect a new root. Therefore, once the root is failure, the RMTS can maintain synchronization accuracy in a long term until a new root is elected.

\section{Conclusions }

A multiple-one-way-broadcast based rapid-flooding time-synchronization protocol is proposed in this paper. The clock skew MLE is employed to generate accurate and fast-converging clock skew estimation, and the MLE clock offset estimation is employed to minimize the delay and to achieve accurate time synchronization. The clock parameters are shared to guarantee the fast convergence rate of large-scale WSNs and to protect against the by-hop error accumulation. Experimental results show that the proposed RMTS converges faster and is more accurate than existing flooding synchronization algorithms.

\section*{Acknowledgment}
The authors would like to thank Zhou Yingyue and Ran Lili from Southwest University of Science and Technology, Mianyang, China for their remarks, and the associate editor and anonymous reviewers for their valuable comments.

\ifCLASSOPTIONcaptionsoff
  \newpage
\fi
{
    \scriptsize
    \bibliographystyle{ieeetr}
    \bibliography{RMTS_TII}
}

\begin{IEEEbiography}[{\includegraphics[width=1in,height=1.25in,clip,keepaspectratio]{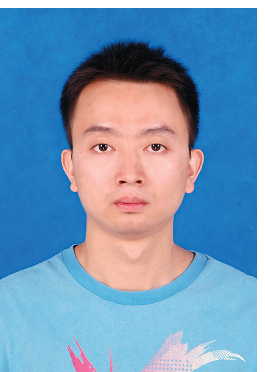}}]{Fanrong Shi}
(S'18) received the B.E. degree in Communication Engineering, the M.E. degree in Communication and Information System, and the  Ph.D. degree in Control Science and Engineering from SWUST, Mianyang, China, in 2009, 2012, 2019 respectively. Currently He is a Lecturer with the School of Information Engineering at the Southwest University of Science and Technology. His research interests include time synchronization and location in wireless sensor networks applications, internet of things, industrial wireless network, wireless synchronous measurement and acquisition, and intelligent instruments.

\end{IEEEbiography}

\begin{IEEEbiography}[{\includegraphics[width=1in,height=1.25in,clip,keepaspectratio]{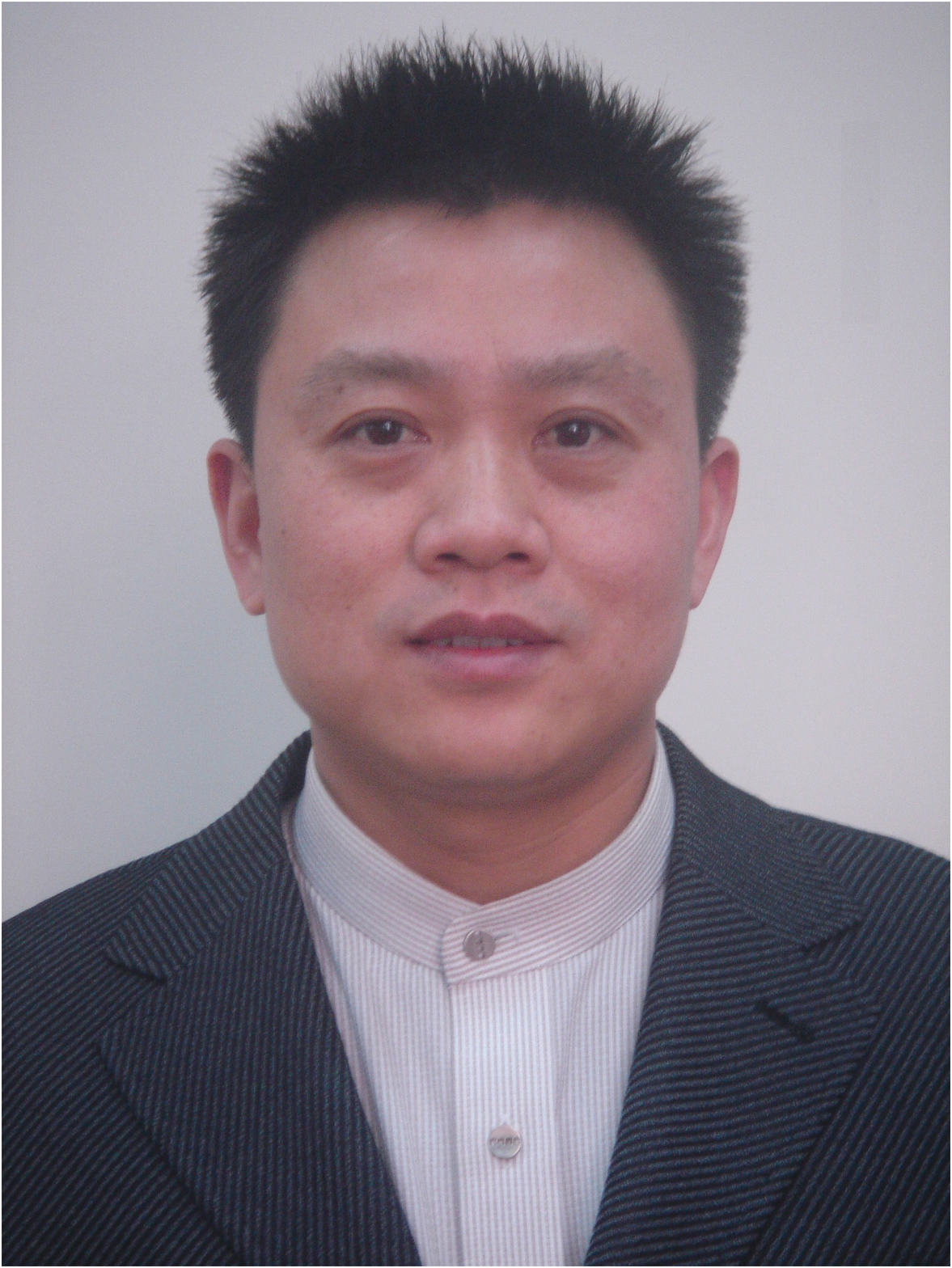}}]{Xianguo Tuo}
received the B.E., M.E. and Ph.D. degrees in Nuclear Geophysics, Environmental Radiation Protection, Applied Nuclear Technology in Geophysics from the Chengdu University of Technology in Chengdu, China, in 1988, 1993, 2001 respectively. From 2006 to 2007, he worked as visiting scholar at the School of Bioscience at University of Nottingham, UK. Since 2001, he is Professor with College of Nuclear Technology and Automation Engineering, Chengdu University of Technology, China. Since 2012, he becomes Professor with the School of National Defense Science and Technology, Southwest University of Science and Technology in Mianyang, Sichuan, China. Currently, he is the Professor and President of Sichuan University of  Science and Engineering, Zigong, Sichuan. Professor Tuo received The National Science Fund for Distinguished Young Scholars in 2011. Currently, his research interests are in detection of radiation, seismic exploration and specialized robots.
\end{IEEEbiography}

\begin{IEEEbiography}[{\includegraphics[width=1in,height=1.25in,clip,keepaspectratio]{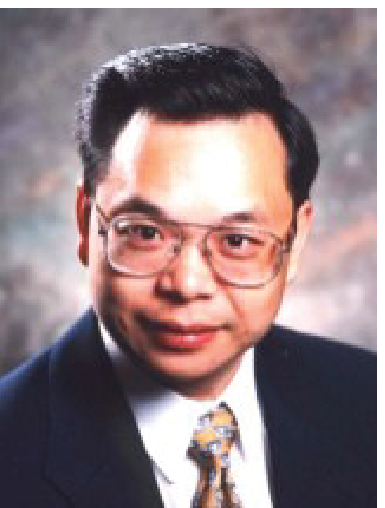}}]{Simon X. Yang}
(S'97, M'99, SM'08) received the B.Sc. degree in engineering physics from Beijing University, China in 1987, the first of two M.Sc.  degrees in biophysics from Chinese Academy of Sciences, Beijing, China in 1990, the second M.Sc. degree in electrical engineering from the University of Houston, USA in 1996, and the Ph.D. degree in electrical and computer engineering from the University of Alberta, Canada in 1999. Currently he is a Professor and the Head of the Advanced Robotics and Intelligent Systems (ARIS) Laboratory at the University of Guelph in Canada. His research interests include intelligent systems, robotics, sensors and multi-sensor fusion, wireless sensor networks, control systems, and computational neuroscience. Prof. Yang serves as the Editor-in-Chief of International Journal of Robotics and Automation, and Associate Editor of IEEE Transactions on Cybernetics, and several other journals. He has involved in the organization of many conferences.
\end{IEEEbiography}

\begin{IEEEbiography}[{\includegraphics[width=1in,clip,keepaspectratio]{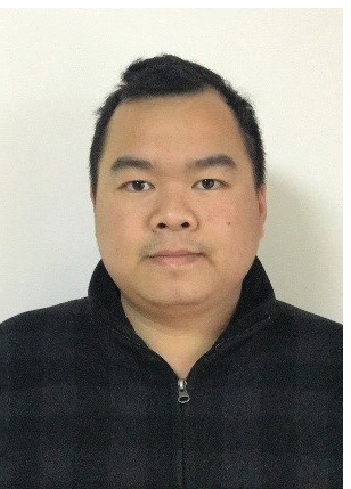}}]{Jing Lu}
is now a Ph.D. student from major of Control Science and Engineering, in Southwest University of science and Technology. In 2014, he finished his graduate project of human-robot interaction subject in LIRMM (Montpellier Laboratory of Informatics, Robotics and Microelectronics), Montpellier and worked in Aldebaran, Paris as an intern student, working at a humanoid robot project. In the same year, he received his M.S degree in Electrical Engineering from Montpellier University Graduate Engineering School. Currently, he focuses on the development of the exploration robot in special environment.
\end{IEEEbiography}

\begin{IEEEbiography}[{\includegraphics[width=1in,height=1.25in,clip,keepaspectratio]{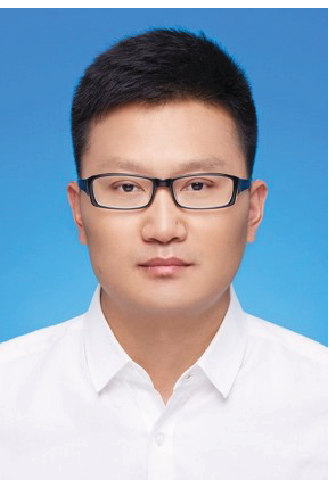}}]{Huailiang Li}
(M'18) received the B.E., M.E., and Ph.D. degrees in signal and information processing from the Chengdu University of Technology, Chengdu, China, in 2007, 2010, and 2013, respectively.
Since 2013, he worked as Associate Professor or Professor with the School of National Defense Science and Technology, Southwest University of Science and Technology in Mianyang, Sichuan, China. He is currently a Professor in the College of Geophysics, Chengdu University of Technology, Chengdu, China. His research interests include data acquisition, signal processing and wireless sensor network in geosciences.
\end{IEEEbiography}

\end{document}